# Automation Impacts on China's Polarized Job Market

Haohui "Caron" Chen[1,2*†], Xun Li[3*†], Morgan Frank[4], Xiaozhen Qin[3], Weipan Xu[3], Manuel Cebrian[4], Iyad Rahwan[4*]

[1]Data61, Commonwealth Scientific and Industrial Research Organisation (CSIRO), Australia.

[2]Faculty of Information Technology, Monash University, Australia.

[3]Sun Yat-sen University, China.

[4]Media Laboratory, Massachusetts Institute of Technology, USA.

*Correspondence to: Xun Li (lixun@mail.sysu.edu.cn), Haohui "Caron" Chen (caronhaohui.chen@data61.csiro.au), and Iyad Rahwan (irahwan@mit.edu).

† These authors contributed equally to this work.

## Abstract

When facing threats from automation, a worker residing in a large Chinese city might not be as lucky as a worker in a large U.S. city, depending on the type of large city in which one resides. Empirical studies found that large U.S. cities exhibit resilience to automation impacts because of the increased occupational and skill specialization. However, in this study, we observe polarized responses in large Chinese cities to automation impacts. The polarization might be attributed to the elaborate master planning of the central government, through which cities are assigned with different industrial goals to achieve globally optimal economic success and, thus, a fast-growing economy. By dividing Chinese cities into two groups based on their administrative levels and premium resources allocated by the central government, we find that Chinese cities follow two distinct industrial development trajectories: one trajectory owning government support leads to a diversified industrial structure and, thus, a diversified job market, and the other leads to specialty cities and, thus, a specialized job market. By revisiting the automation impacts on a polarized job market, we observe a Simpson's paradox through which a larger city of a diversified job market results in greater resilience, whereas larger cities of specialized job markets are more susceptible. These findings inform policy makers to deploy appropriate policies to mitigate the polarized automation impacts.

**One Sentence Summary:** China's top-down, centrally planned specialization of cities makes large Chinese cities less resilient to impact from automation technologies.

# 1 Introduction

Advances in artificial intelligence (AI) and robotics technologies have revived the concerns of technological unemployment. Frey and Osborne (*1*) estimated that more than 47% of U.S. jobs were at high risk of computerization, and an alternative OECD study (*2*) found a more modest 9% of U.S. jobs were at risk. Based on the method in (*1*), Frey et al. (*3*) expected the impacts on China's job market to be devastating and accounted for 77% of employment at risk, or approximately 550 million jobs (*4*). Even though these figures are highly disputed, they foresee unavoidable impacts on the global job market for the upcoming decades. China is at the precipice of becoming a developed economy, and this stage could reflect the greatest income inequality (*5*) and resource accessibility (*6*). The impacts on the job market would certainly exacerbate the inequality, and the resulting destabilization would be a significant problem for not just China but also the world.

The expected impacts on China's job market are significantly greater than those on OECD countries (*3*). This result could stem from differences in economic structures. China has become the world's factory, receiving manufacturing orders from around the world; at the same time, increasing domestic demand has elevated its manufacturing capacities. However, jobs in the manufacturing sector are among the most susceptible to automation (*1*). Moreover, to secure food safety for almost 1.4 billion people, a significant labor force is in the farming sectors, occupations of which are also susceptible to automation. Both aggravate the overall impact on China's job market.

The rapid urbanization that has developed megacities, such as Beijing, Shanghai, and Shenzhen, has allowed China to remedy the economic gaps to the developed world. Whereas China's cities accommodate more than half its population (*4*), the impacts from automation at the city level in China have never been studied. Frank et al. (*7*) found that large U.S. cities are more resilient to the impacts of automation because increasing numbers of specialized occupations and skills that are resilient to automation were observed, along with the growing size. Given that China's cities also express similar scaling behavior in productivity and R&D employment (*8*), we also expect the resilience of China's large cities. We estimate job impact rates for 102 cities in China based on the method in (*1*) (see the Results for details) and found polarized responses along with the growing size; some large cities are also estimated to be susceptible to automation (Fig. 1). For example, Nanyang, the fifth largest city in China, is famous for its farming produce, has 79% employment in the primary industry and is estimated to have 83% jobs at risk. These "specialty cities" in which only one or a few industries dominate, could be explained as the legacy of China's planned economy, and the central government assigned specific development targets to different regions (*9*). After the economic reform, regional governments gain greater autonomy in choosing industrial strategies. For example, between 2016 and 2017, the central government approved 403 local government applications to initiate "specialty towns"—focused on one particular industry—and aims to have 1,000 of them by 2020 (*10*). However, given the regional endowments and the central government's preferences, China's cities have developed a polarized industrial structure over increasing city size, for which some regions adapted a more diversified industrial structure, similar to U.S. cities (*11*), but some adapted a path-dependent development trajectory (*12*) to become "specialty cities". Given industrial polarization, China's cities develop a polarized job market. Therefore, this study aims to understand the emerging impacts from automation technologies on China's polarized job market.

## 2 Materials and Methods

### 2.1 Estimation of job impacts in China

To utilize the estimated job impact rate of (1), we build a correspondence table between China's Grand Classification of Occupations (GCO) and the U.S.'s Standard Occupational Classification (SOC). There are 8 chapters, 66 sections, and 413 subsections of occupations in the GCO. The GCO's 413 subsectional occupations are mapped to one or multiple occupations from 702 SOC 6-digits occupations based on the titles and corresponding descriptions. The matching process is as follows.

(1) Three paid students ($N = 3$) with backgrounds in human geography mapped GCO subsectional occupations ($J = 413$) to as many SOC 6-digits occupations ($K = 702$) as appropriate based on titles and corresponding descriptions, respectively. This step generates an array of mappings for individual GCO occupations: $occupation_{GCO}(j) \in R^{702}$, $j \in J$, where $R$ is a vector with element $occupation_{SOC}(k) \in \{0, 1, 2, 3\}, k \in K$. The value of $occupation_{SOC}$ indicates the number of occurrences of such an occupation as mapped by students.

(2) If a GCO occupation has all candidates $occupation_{SOC} \leq 1$, the authors decided on the SOC to which that occupation should be mapped, the corresponding $occupation_{SOC} \leftarrow 1$, and the rest $occupation_{SOC} \leftarrow 0$.

(3) The commonly matched SOC occupations ($occupation_{SOC} \geq 2$) are kept and the corresponding value is changed $occupation_{SOC} \leftarrow 1$, and the rest $occupation_{SOC} \leftarrow 0$. The final output of the correspondence table is a matrix $R^{413 \times 702}$, in which each element is $occupation_{SOC} \in \{0,1\}$.

We cannot find a mapping for occupations in the "leader of political or state-owned entities" SOC sector; therefore, we assign a zero impact rate for them given the special nature of this type of occupation. Subsequently, we compute the mean impact rate of all $occupation_{SOC} = 1$ and assign the mean value as the computerization probability to the corresponding GCO occupations.

The Sixth National Population Census (hereafter referred to as the Census) carried out by the National Bureau of Statistics of China in 2010 identifies the employment distribution of 413 subsectional occupations and 95 industrial subsectors across 295 cities (metropolitan areas). However, only 102 city governments make this Census available in paper form, which we converted into electronic form.

Industry, job, and impact rate: industry and job diversity are computed using normalized Shannon entropy (29), as follows. Normalized Shannon entropy has been used in urban science (7, 30).

$$H_{job}(m) = -\sum_{j \in Jobs_m}[p_m(j) \times log(p_m(j))]/log(Jobs_m) \qquad [2]$$

$$H_{industry}(m) = -\sum_{k \in Industries_m}[p_m(k) \times log(p_m(k))]/log(Industries_m) \qquad [3]$$

$p_m(j)$ denotes the probability of a worker in city $m$ having job $j$, $p_m(k)$ denotes the probability of a worker in city $m$ working in industry $k$, and $Industries_m$ is the set of industry sectors in city $m$. In the computation of job diversity $H_{job}(m)$, we use $Jobs_m \in 413\ sectional\ jobs$, and in the computation of industry diversity $H_{industry}(m)$, we use $Industries_m \in 95\ industry\ sectors$.

To estimate the relationships among city size, job diversity, industry diversity, distance to the closest elite cities, net population gains, fixed asset investments, and impact rate, we build linear regression models and estimate the coefficients using the ordinary least squares (OLS) method. The response variable and predictor can be one of the previously listed variables, and the predictor can be log-transformed. If the predictor is log-transformed, it is explicitly addressed in the caption of the corresponding figure. $\beta$ reports the coefficient of the predator. To estimate the scaling behavior of occupation growth, we build a linear regression with a log-log transformation. $\beta$ reports the scaling exponents.

## 2.2 China's higher education system and bullet trains

By 2010, 39 and 109 universities have been funded by projects "985" and "211," respectively (some are funded by both concurrently). The data are available from the official website of the Ministry of Education of China: http://www.moe.gov.cn/srcsite/A03/moe_634/.

Another significant investment devoted to Chinese cities is the bullet trains railway network. The daily operating frequency of bullet trains in each city can be found at the online bullet train ticket office: http://www.gaotie.cn/. Both university and bullet train statistics are available in Table. S1.

## 2.3 Occupation space

We adapt the methodology from (22) to construct the occupation space.

Revealed comparative advantage (RCA): $RCA_{m,j}$ measures, for city $m$, occupation $j$'s relative level to the national average. $x_{m,j}$ is the total number of jobs of occupation $j$ in city $m$. When $RCA_{m,j} > 1$, city $m$ has more jobs of occupation $j$ as a share of its total numbers of jobs than the national average, and vice versa. An occupation with $RCA_{m,j} > 1$ is considered one of the city's advantaged occupations.

$$RCA_{m,j} = (x_{m,j}/Jobs_m)/(Jobs_m/\sum_m Jobs_m) \qquad [4]$$

Proximity: The proximity between two occupations $i$ and $j$, denoted as $\phi_{i,j}$, is the minimum of the pairwise conditional probabilities of a city accommodates an occupation given that it also accommodates another. A closer proximity indicates a higher correlation between two occupations and a closer spacing of the two occupation nodes in the occupation space (Fig. 4a). The value of $\phi_{i,j}$ is between 0 and 1.

$$\phi_{i,j} = min\{Pr(RCA_{m,i} \geq 1|RCA_{m,j} \geq 1), Pr(RCA_{m,j} \geq 1|RCA_{m,i} \geq 1)\}$$

[5]

National occupation space. First, we construct a 413 * 413 occupation proximity matrix. Second, we generate a maximum spanning tree (MST) based on the proximity matrix to generate the network's skeleton and complement it with additional proximity links whose $\phi_{i,j} > 0.66$. For better visualization, we use a force spring layout to visualize the network.

## 3 Result

### 3.1 The impacts of automation

We estimated the impacts of automation on China's job market at the city level using [1] as adapted from (*7*).

$$E_m = \frac{\sum_{j \in Jobs_m} p_{auto}(j) \cdot f_m(j)}{\sum_{j \in Jobs} f_m(j)} \quad [1]$$

$E_m$ is the expected job impact rate for city $m$, $f_m(j)$ denotes the number of workers in city $m$ with job $j$, and $Jobs_m$ is the set of job types in city $m$. $p_{auto}(j)$, the probability of computerization for job $j$ is adopted from (*1*). On average, 79% of jobs are expected to be at high risk for 102 cities in China, which is close to the estimation of (*3*) (see Table. S1 for estimations for all 102 cities). Well-known large cities, such as Beijing, Shanghai, Guangzhou, and Shenzhen, exhibit resilience to automation technologies, with expected impact rates of 64%, 67%, 69%, and 72%, respectively, whereas other large cities such as Zhumadian and Nanyang are the opposite, both with job impacts as high as 83% (Fig. 1). Different from the findings of (*7*), city size does not express strong correlations over job impacts as a whole.

Susceptible large cities have long been regarded as "specialty cities", which are either specialized in farming (e.g., Nanyang), mining (e.g., Pingdingshan), or manufacturing (e.g., Taizhou), whereas resilient large cities are either innovation hubs (e.g., Beijing) or financial (e.g., Shanghai) or regional services centers (e.g., Guangzhou). The former industrial sectors tend to employ jobs of higher computerization risks, whereas the opposite is true for the latter. The underlying industrial structure might constitute distinct job market and, as a result, cause Chinese cities' polarized responses to automation impacts. Therefore, it is worth studying the relationships between each of them.

## 3.2 Polarized job market

To understand China's industrial structure, we perform a principal component analysis (PCA) (*13*) on the employment distributions across 20 industrial sectors (see Fig. 2). PCA has been used in urban science (*14*, *15*). The PCA result shows that two leading principal components (PCs) accumulate more than 85% of the total variances. Therefore, by analyzing two PCs' industrial employment compositions and cities' PC scores, one can address China's industrial structure. A group of vectors consisting of tertiary industries pointing in the 9 o'clock direction on PC1 corresponds to the coexistence of the tertiary industries in the same set of cities. However, farming and mining both have minor contributions regarding PC1 but significant regarding PC2, indicating that both industries tend to co-locate but dislocate to the tertiary industries. Manufacturing contributes to both PCs but affects inversely against farming and mining regarding PC2. Thus, from the composition of PCs, we can claim two PCs as a servicing index and a natural resources index, respectively. Cities' positions on this plot unveil their industrial structure. For example, Beijing and Shanghai have significant servicing but minor natural resources' indexing scores, addressing their advantages in servicing roles in the Chinese economy. In this plot, two types of large cities clearly emerge. One type is servicing centers and the other type is natural resources centers, which also correspond to two types of large cities that face polarized automation impacts. Interestingly, if we set one of the smallest cities in this study, Sanya, as the starting point, we can observe that cities tend to follow two trajectories toward those two extreme industrial structures, along with increasing size (dashed vectors in Fig. 2). However, to the best of our knowledge, no study has addressed Chinese cities' polarized industrial path-taking behavior; therefore, this study also attempts to understand its underlying driving forces. The color codes of cities in Fig. 2 will be subsequently addressed.

The distinct industrial trajectories could have been elaborately planned for decades by the central government to achieve global optima of economic success. However, given the lack of historical planning materials, we cannot investigate how cities' or regions' missions were assigned. Instead, we can infer the central government's master plan toward each city by studying the spatial allocations of premium resources or, to be more specific, the research facilities and transport infrastructure. To develop innovation capabilities and international competitiveness, China launched two national education projects—"985" and "211"—in the 1990s to enable a limited number of universities to become world-class research facilities (*16*). As a result, these premium research facilities have significantly improved the innovation competitiveness of corresponding cities (*17*). By 2010, 113 universities have been funded by the "985" and "211" projects.

Another significant premium resource is the bullet train railway network invested in and managed by the central government. The allocation of bullet train stations and the operating frequency address the importance of a corresponding city as the regional service center, such as a freight forwarder. Moreover, bullet trains also exhibited strong effects in the overall competitiveness of a city (*18*). In this study, we use daily operating frequency of the bullet train as a proxy for infrastructure investments by the central government (see Materials and Methods for more details about the Chinese higher education system and bullet train distributions). Using k-means clustering ($k = 2$) on those two features, two groups of cities were found, and we name them premium ($N_{premium} = 20$) and non-premium cities ($N_{non-prem} = 82$).

It is not a complete surprise that the majority of premium cities are direct-controlled municipalities, sub-provincial level municipalities, or provincial capitals. These cities enjoy administrative powers over their peers. One of the exclusive advantages is direct communications with the central government; therefore, state-owned and multinational companies prefer to reside in them to reduce communication costs and to stay informed of volatile policies and regulations. Fixed asset investments can be a good proxy for addressing the preferences of those companies. Fig. S1 shows that fixed asset investments grow linearly with city size in premium cities and cities of higher administrative power, whereas they grow sub-linearly in non-premium cities and cities of lower administrative power. The administrative levels of each city were successively imposed by the central government after the 1950s until the 1990s. Therefore, because this long-lasting administrative division might also unveil the master planning of the government over the past decades, we also group cities by their administrative levels and call them elite and non-elite cities. In this study, there are a total of 19 elite and 83 non-elite cities. Among them, 16 elite and 4 non-elite cities appear on the list of premium cities (see Table. S1 for the full list of both division systems).

In Fig. 1b and Fig. 1c, we find that both large premium and elite cities exhibit resilience to automation impacts relative to the others. Moreover, we observe a Simpson's paradox (*19*), for which the larger cities on the advantaged side (premium and elite) have greater resilience to automation impacts; additionally, the larger cities on the non-advantaged side are more susceptible. In this paradox, small non-advantaged cities (non-premium and non-elite) are resilient to automation impacts, which is opposite of the findings for the United States (*7*). We also compare city size effects between Chinese and U.S. cities (see Fig. S2). Even though the overall impact in China is significantly higher than in the United States, Chinese cities on the advantaged sides exhibit stronger size effects than U.S. cities ($|\beta_{premium}| > |\beta_{elite}| > |\beta_{US}|$). Therefore, cities enjoying more premium resources or higher administrative power help mitigate the automation impacts to a great extent.

In Fig. 2, along with the two distinct industrial trajectories that we hypothesized, premium cities tend to develop a more diversified industrial structure along with increasing size, which is similar to U.S. cities (*11*), whereas non-premium cities are adapted to be specialized in either farming and mining or manufacturing. Even though we cannot investigate how cities are assigned with specific missions given the lack of historical planning materials, we can infer from the previous finding that the central government has put in force two distinct industrial strategies over these two types of cities. Given that China started its reform as a poor country, to achieve a fast-growing economy, limited premium resources can only be deployed in a few trailblazer cities and become innovation hubs and financial and regional services centers. The other cities might need to specialize in a few industries to benefit from the scale economies (*20*). Thus, global optima of economic success can be achieved. Moreover, we also find that non-advantaged cities tend to develop a more diversified industrial structure when they are geographically close to elite cities (Fig. S3). This might aggravate polarization because both innovations and diversified job markets exist in elite cities (see Fig. 3a). Skilled and well-educated workers tend to migrate to these cities to earn higher wages (*21*) and, as a result, non-advantaged cities farther from elite cities have lower chances of attracting those workers. Therefore, because "specialty cities" might lack skilled and well-educated workers to develop high-tech industries, they continue to specialize in low-tech industries, such as farming and mining. We confirm the population loss of non-advantaged cities along with the growing distance to the closest elite cities (Fig. S4). We use the distance to the closest elite cities instead of to the premium cities because elite cities are more geographically distributed than premium cities (mainly located in eastern China), and non-premium cities in western China could benefit more from their elite neighbors than distant premium cities in eastern China. Moreover, the distance computation involves not just the elite cities in this study but also those beyond this study given the lack of detailed census records.

Because a diversified industrial structure constitutes a diversified job market and vice versa (strong correlation between both are observed in Fig. S5), polarized job diversities of two groups of cities over city size is also observed (Fig. 3a). There are some large non-premium cities also developing diversified job market, such as Taizhou. Those cities might have been taking similar industrial trajectories as those of premium cities, given that they are closer than their peers to diversified elite cities. More importantly, we find that job diversity significantly affects expected job impacts (Fig. 3b). Similar to findings in the United States (*7*), a diversified job market can mitigate automation impacts to a great extent.

By studying occupation growth over city size, we can understand how job market composition affects the expected job impact rate (see Figs. 3c & 3d, and see Fig. S6 for results under the administrative division). The distinct occupation growth patterns of premium and non-premium cities show that, in non-premium cities, susceptible occupations (e.g., primary industry, production, and construction) grow super-linearly with city size. In contrast, they grow either sub-linearly, linearly, or even negatively with city size in premium cities. The most resilient occupations, such as public services, quality checks and measurements, professionals, and film and music, are among the most distinct scaling exponents between premium and non-premium cities.

### 3.3 Evolution of occupational structure

The product space of Hidalgo et al. (*22*) offers a compelling illustration that addresses the relationships between world trade products and the roles played by different countries. In the product space, products are connected based on their co-location probability, and the core area consists of sophisticated products such as metal products, machinery, and chemicals, whereas the periphery consists of fishing, tropical, and cereal agriculture. Industrialized countries are found to be dominant in exporting products in the core area, whereas non-industrialized ones are dominant in the periphery.

Also, empirically proven was that one developing country can traverse through links of the network to the core area to gain relative advantages with some sophisticated products, which can later constitute the base of further traversing and, thus, industrialization. Inspired by the product space, we hypothesize that advantaged cities grow to be hubs of innovation, finance, arts, and services, and become service centers for surrounding non-advantaged cities that develop into "specialty cities" given the master planning of the central government. Therefore, we construct the first occupation space for China based on the co-location probability of any two occupations (Fig. 4 and see Materials and Methods for the detailed process).

Similar to the occupation space of the United States (*23*), China's occupation space also has a service sector and professional occupations in the core area and production and farming at the periphery (Fig. 4a). Moreover, occupations at the core express resilience to the impacts of automation, whereas occupations at the periphery tend to be susceptible to automation (Fig. 4b).

To illustrate the dominance of cities in certain parts of the occupation space, we highlight a few symbolic cities based on their relative advantages. Beijing (Fig. 5b) is dominant in the professional sector at the core, whereas the small premium city of Putian (Fig. 5a), known as the "shoes-making city", is dominant in the production sector at the periphery. The large non-premium city Nanyang (Fig. 5e) is dominant in 28 occupations, mainly in farming and production at the periphery, whereas the small, non-premium city of Sanya (Fig. 5d), known as the tourist city, is dominant in the service sector at the core. We observe two distinct evolution paths of the job market for both premium and non-premium cities (Fig. 5c): premium cities transit from the periphery to the core and the non-premium cities transit in the opposite direction. This confirms that premium cities grow to be services centers, whereas non-premium cities grow to become "specialty cities". Polarization might achieve economic optima for China as a whole; however, in the context of technological unemployment, it might cause significant threats to non-advantaged cities and inequality for the entire job market.

## 4 Limitations

We can only access the census data for 102 cities, which were made available to the public by the local governments. Even though they are geographically distributed and have different sizes and GDPs, they are just above one-third of the entire population, that is, 295 cities. In this regard, audiences should be cautious in using the conclusions drawn from this limited number of city samples.

Similar to (*7*), many limitations inherent in occupation-level estimations apply to this study as well. Moreover, China lacks statistics on skill and task distributions across occupations, and we cannot perform the task-based approach (*2*) in estimating the jobs impact in this study. The job impacts in China could be significantly weaker than the estimation by (*3*) and this study. Therefore, policy makers are encouraged to interpret the impact rates as relative values for comparing the impacts between cities.

# 5 Discussion

The automation impacts on the job market has never been as tensional as it is today. The main concern is that the loss of jobs to automation could outstrip the demand created by the corresponding increasing productivity (*24*). Appropriate policies should be deployed to help mitigate rapid changes to the job market. However, an understanding of the emerging changes at the city level has never been studied for the most populous country, China. This study attempts to compute the job impacts for 112 Chinese cities and addresses the reasons behind the polarized responses over increasing city size.

Given a lack of access to historical planning materials, we cannot investigate how missions were assigned to each city. However, by grouping cities based on their administrative level and allocation of premium resources, we find two distinct industrial development trajectories among them that unveil the central government's master planning regarding the pursuit of the global optima of economic success. China's polarized industrial structure has constituted a polarized job market in which susceptible occupations appear more in non-advantaged cities and resilient occupations strive in advantaged cities and, accordingly, polarize responses to the emerging automation impacts. Moreover, non-advantaged cities can benefit significantly from their neighbor advantaged cities of a diversified industrial structure, which might further exaggerate the polarization of automation impacts. In all, sufficient grounds exist to believe that cities' lack of administrative power or premium resources and distance from advantaged cities could be left behind in the second machine age (*24*).

Up-skills would be one of the most feasible ways to mitigate automation impacts (*25*) but involve appropriate allocation of educational resources, especially for lifelong learning. We find that China's existing allocation of educational resources might not match where they are needed the most (Figs. S7 & S8). The growth of vocational teachers and vocational schools are linearly and sub-linearly correlated with size in non-advantaged cities, respectively. However, because large, non-advantaged cities could suffer the most from automation impacts, vocational education facilities should at least super-linearly grow to the city size in non-advantaged cities. Thus, the central government should play an important role in motivating or subsidizing appropriate policies to support vocational education where it is most needed.

To the best of our knowledge, this study is the first to explore the city division system in China and to find the polarization of industrial structure and the job market. The city division system could change urban research in planned economies (e.g., scaling laws (*8*) and agglomeration economies (*26*)) given that non-advantaged cities have not been following organic growth, whereas advantaged cities enjoy the fruits of their non-advantaged peers. Empirical studies tend to investigate cities as an entire population and might overlook the fundamental differences between organic and planned economies. For example, in (*26*), China was found to exhibit smaller estimates of urban agglomeration elasticities than other countries. However, we might expect a different finding when Chinese cities are treated in a two-study population. Two distinct urban scaling exponents were found between eastern and western Europe (*27*); therefore, we believe that it is worth revisiting scaling laws in China using this city division.

Indeed, journalists and even expert commentators have successfully portrayed technological unemployment as an unsettling picture of the future of work. However, most overlook the complementarities between automation and labor. As Autor (*28*) addressed, automation does substitute tasks but also increases productivity and earnings and, as a result, augments higher demand for labor. In China, the future research about the complementarities should take into account the polarized job market.

# References

1. C. B. Frey, M. A. Osborne, The future of employment: How susceptible are jobs to computerisation? Technol. Forecast. Soc. Change. 114, 254–280 (2017).

2. M. Arntz, T. Gregory, U. Zierahn, The Risk of Automation for Jobs in OECD Countries: A COMPARATIVE ANALYSIS. OECD Social, Employment, and Migration Working Papers; Paris (2016), pp. 0_1,3–5,7–34.

3. C. B. Frey, M. A. Osborne, C. Holmes, E. Rahbari, Technology at work v2. 0: the future is not what it used to be. Citi GPS: Global (2016) (available at http://www.voced.edu.au/content/ngv:76273).

4. National Bureau of Statistics of China, Sixth National Population Census of the People's Republic of China (2011), (available at http://www.stats.gov.cn/tjsj/pcsj/rkpc/6rp/indexch.htm).

5. Y. Xie, X. Zhou, Income inequality in today's China. Proc. Natl. Acad. Sci. U. S. A. 111, 6928–6933 (2014).

6. C. Brelsford, J. Lobo, J. Hand, L. M. A. Bettencourt, Heterogeneity and scale of sustainable development in cities. Proc. Natl. Acad. Sci. U. S. A. 114, 8963–8968 (2017).

7. M. R. Frank, L. Sun, M. Cebrian, H. Youn, I. Rahwan, Small cities face greater impact from automation. J. R. Soc. Interface. 15 (2018), doi:10.1098/rsif.2017.0946.

8. L. M. A. Bettencourt, J. Lobo, D. Helbing, C. Kühnert, G. B. West, Growth, innovation, scaling, and the pace of life in cities. Proc. Natl. Acad. Sci. U. S. A. 104, 7301–7306 (2007).

9. A. Gar-on Yeh, F. Wu, The transformation of the urban planning system in China from a centrally-planned to transitional economy. Prog. Plann. 51, 167–252 (1999).

10. T. Economist, China pushes towns to brand themselves, then regrets it. The Economist (2017), (available at https://www.economist.com/news/china/21732826-officials-beijing-fret-local-boosters-are-getting-carried-away-china-pushes-towns-brand).

11. G. Duranton, D. Puga, Diversity and Specialisation in Cities: Why, Where and When Does it Matter? Urban Stud. 37, 533–555 (2000).

12. S. Zhu, C. He, Y. Zhou, How to jump further and catch up? Path-breaking in an uneven industry space. J. Econ. Geogr., lbw047 (2017).

13. I. Jolliffe, in International Encyclopedia of Statistical Science, M. Lovric, Ed. (Springer Berlin Heidelberg, Berlin, Heidelberg, 2011), pp. 1094–1096.

14. A. Chiesura, The role of urban parks for the sustainable city. Landsc. Urban Plan. 68, 129–138 (2004).

15. J. Zhu, Data envelopment analysis vs. principal component analysis: An illustrative study of economic performance of Chinese cities. Eur. J. Oper. Res. 111, 50–61 (1998).

16. L. Lixu, China's higher education reform 1998–2003: A summary. Asia Pacific Education Review. 5, 14–22 (2004).

17. Y. Jiang, J. Shen, Measuring the urban competitiveness of Chinese cities in 2000. Cities. 27, 307–314 (2010).

18. S. Zheng, M. E. Kahn, China's bullet trains facilitate market integration and mitigate the cost of megacity growth. Proceedings of the National Academy of Sciences. 110, E1248–E1253 (2013).

19. E. H. Simpson, The Interpretation of Interaction in Contingency Tables. J. R. Stat. Soc. Series B Stat. Methodol. 13, 238–241 (1951).

20. P. Krugman, Scale Economies, Product Differentiation, and the Pattern of Trade. Am. Econ. Rev. 70, 950–959 (1980).

21. J. Bessen, Learning by Doing: The Real Connection Between Innovation, Wages, and Wealth (Yale University Press, 2015).

22. C. A. Hidalgo, B. Klinger, A.-L. Barabási, R. Hausmann, The product space conditions the development of nations. Science. 317, 482–487 (2007).

23. R. Muneepeerakul, J. Lobo, S. T. Shutters, A. Goméz-Liévano, M. R. Qubbaj, Urban economies and occupation space: can they get "there" from "here"? PLoS One. 8, e73676 (2013).

24. E. Brynjolfsson, A. McAfee, The Second Machine Age: Work, Progress, and Prosperity in a Time of Brilliant Technologies (W. W. Norton & Company, 2014).

25. A. Spitz-Oener, Technical Change, Job Tasks, and Rising Educational Demands: Looking outside the Wage Structure. J. Labor Econ. 24, 235–270 (2006).

26. P. C. Melo, D. J. Graham, R. B. Noland, A meta-analysis of estimates of urban agglomeration economies. Reg. Sci. Urban Econ. 39, 332–342 (2009).

27. E. Strano, V. Sood, Rich and Poor Cities in Europe. An Urban Scaling Approach to Mapping the European Economic Transition. PLoS One. 11, e0159465 (2016).

28. D. H. Autor, Why Are There Still So Many Jobs? The History and Future of Workplace Automation. J. Econ. Perspect. 29, 3–30 (2015).

29. U. Kumar, V. Kumar, J. N. Kapur, NORMALIZED MEASURES OF ENTROPY. Int. J. Gen. Syst. 12, 55–69 (1986).

30. N. Eagle, M. Macy, R. Claxton, Network Diversity and Economic Development. Science. 328, 1029–1031 (2010).

31. C. Peng et al., Environment. Building a "green" railway in China. Science. 316, 546–547 (2007).

**Acknowledgments:** We thank Esteban Moro (Universidad Carlos III de Madrid) and Lijun Sun (McGill University) for helpful suggestions and comments on this study. We also thank the anonymous reviewers for their valuable suggestions.

**Author contributions:** X.L., H.C., M.C. and I.R. conceived of the research question. X.L., P.X.   and X.Q. processed the Census data. X.L., H.C., M.F. and M.C. interpreted the result. X.L., H.C., M.F., X.Q., M.C. and I.R drafted the manuscript and compiled supplementary information. All authors edited the manuscript and supplementary information and aided in concept development.

**Data and materials availability:** All data needed to evaluate the conclusions in the paper are present in the paper and/or the Supplementary Materials. Additional data used in our study, including raw Census data and city statistics, will be Publicly available for research purpose.

**Fig. 1. Expected job impact rate over city size.** (a) Geographical distributions of expected job impact from automation across China's cities. Premium cities are marked with red stars. (b) and (c) Expected job impact rate over city size. In (b), cities are grouped using k-means clustering according to the educational resources and transport infrastructures funded by the central government. In (c), cities are grouped based on their administrative levels. We build linear regression models using $log_{10}(city\ size)$ as instrumental variables and job impact rate as responses. The estimations of the size effects are presented in the inset table. In the model, city size is log-transformed and does not correlate with job impacts as a whole ($p_{value} = 0.86$), but does within groups.

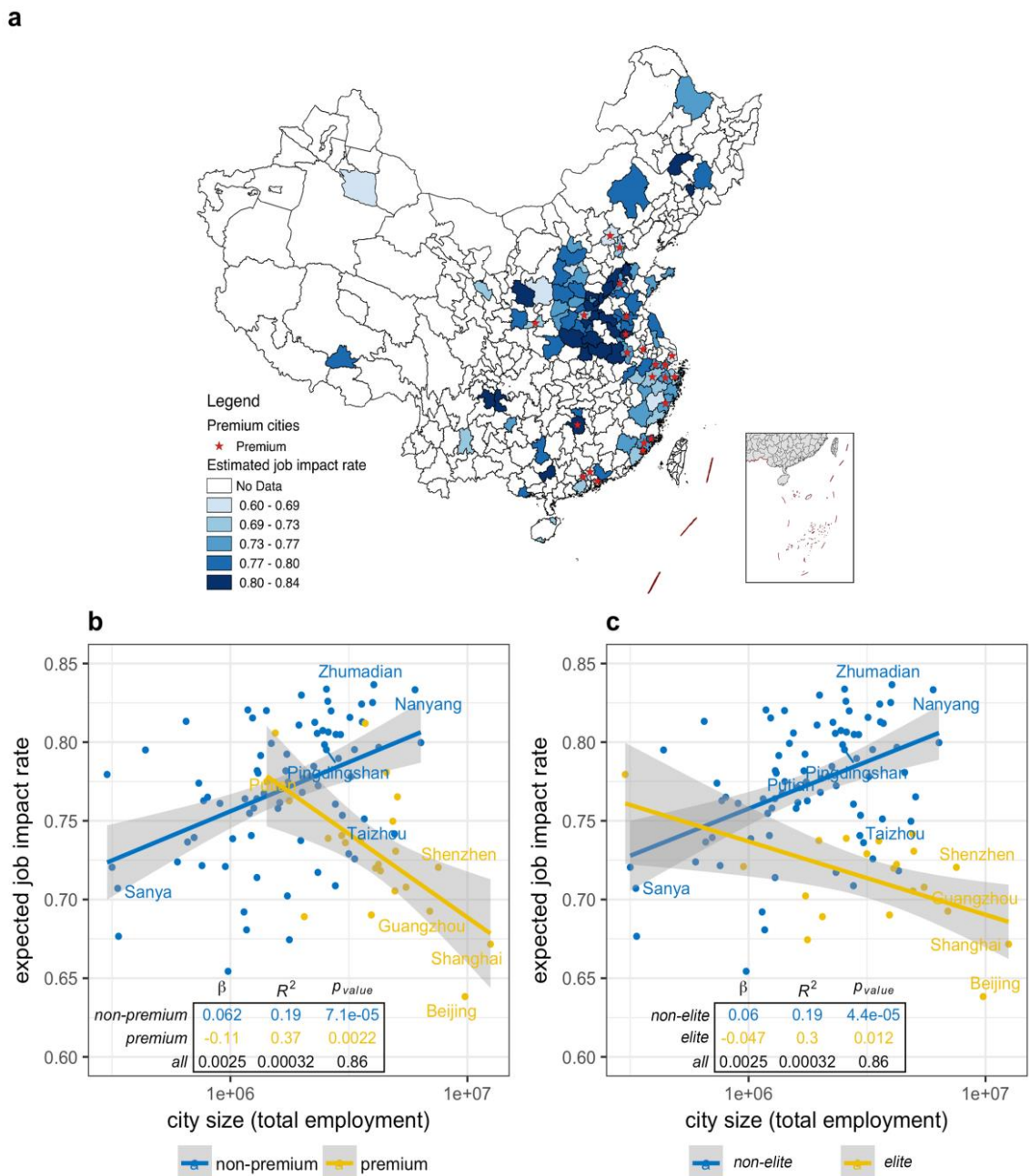

**Fig. 2. China's industrial structure obtained from PCA on the employment distributions across 20 industry sectors.** The solid vectors represent the coefficients of industrial employments on the PCs, the nodes represent cities' corresponding PC

scores, and the dashed vectors starting from Sanya indicate two arbitrary directions that illustrate how small cities transform its industrial structure along with increasing size. The size of the nodes is proportional to the city size. Two leading PCs explain 78.7% and 6.3% of the overall variances, respectively. Two eclipses show the 95% confidence interval (CI) of PC scores of corresponding cities.

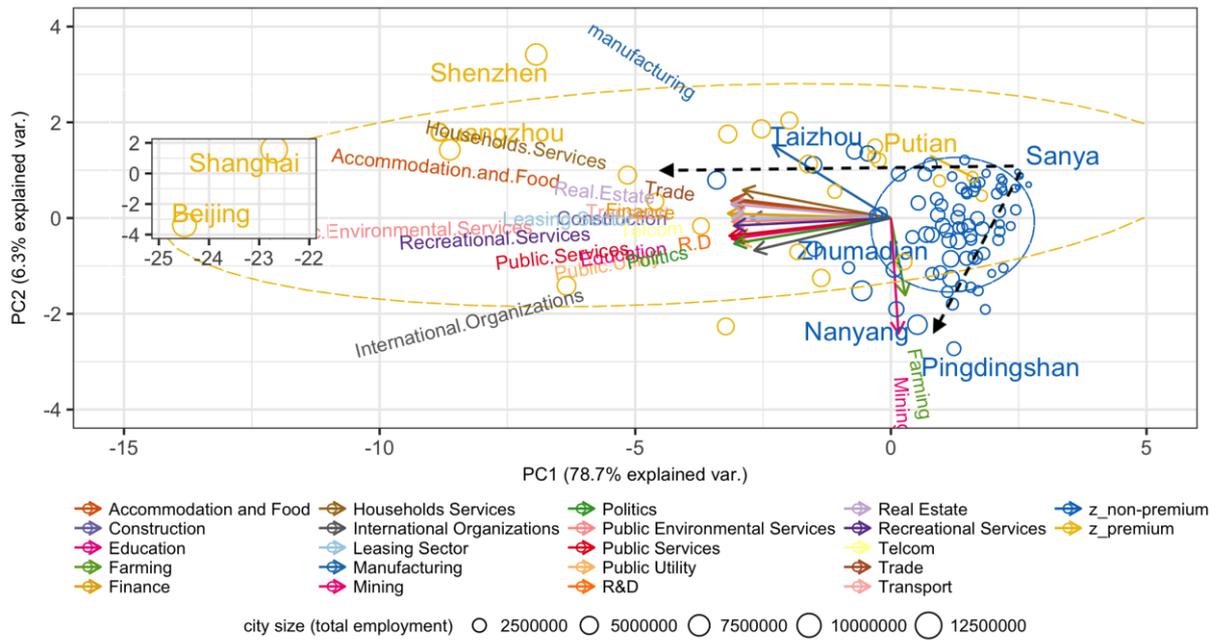

**Fig. 3. Job diversity, impact rate, and occupation growth between premium and non-premium cities.** (a) Job diversity over city size. The node size is proportional to the distance to the closest elite cities. (b) Expected job impact rates over job diversity. The node size is proportional to the city size. Linear regression results are shown in the tables. City size is log-transformed in the model. Panels (c) and (d) show occupation growth over city size in non-premium and premium cities, respectively. Points are vertically shifted according to linear fit in a log scale and the black dashed line has a slope of 1 for reference.

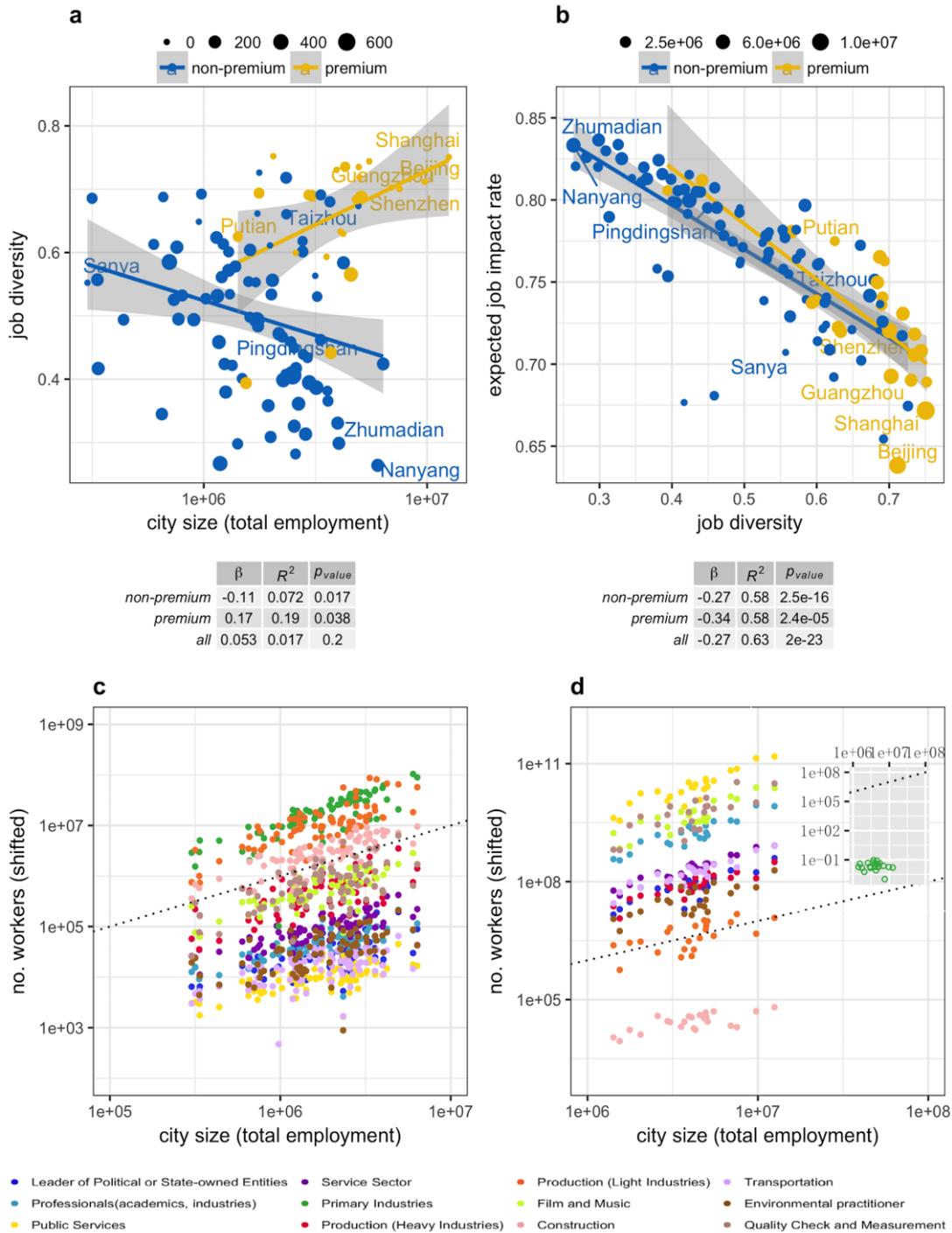

**Fig. 4. Occupation space across 102 Chinese cities.** Panel (a) shows the occupation space; panel (b) shows the relationship between an occupation's closeness centrality and automation rate. The dashed line indicates the best fit of the linear regression, which reports a negative relationship between both.

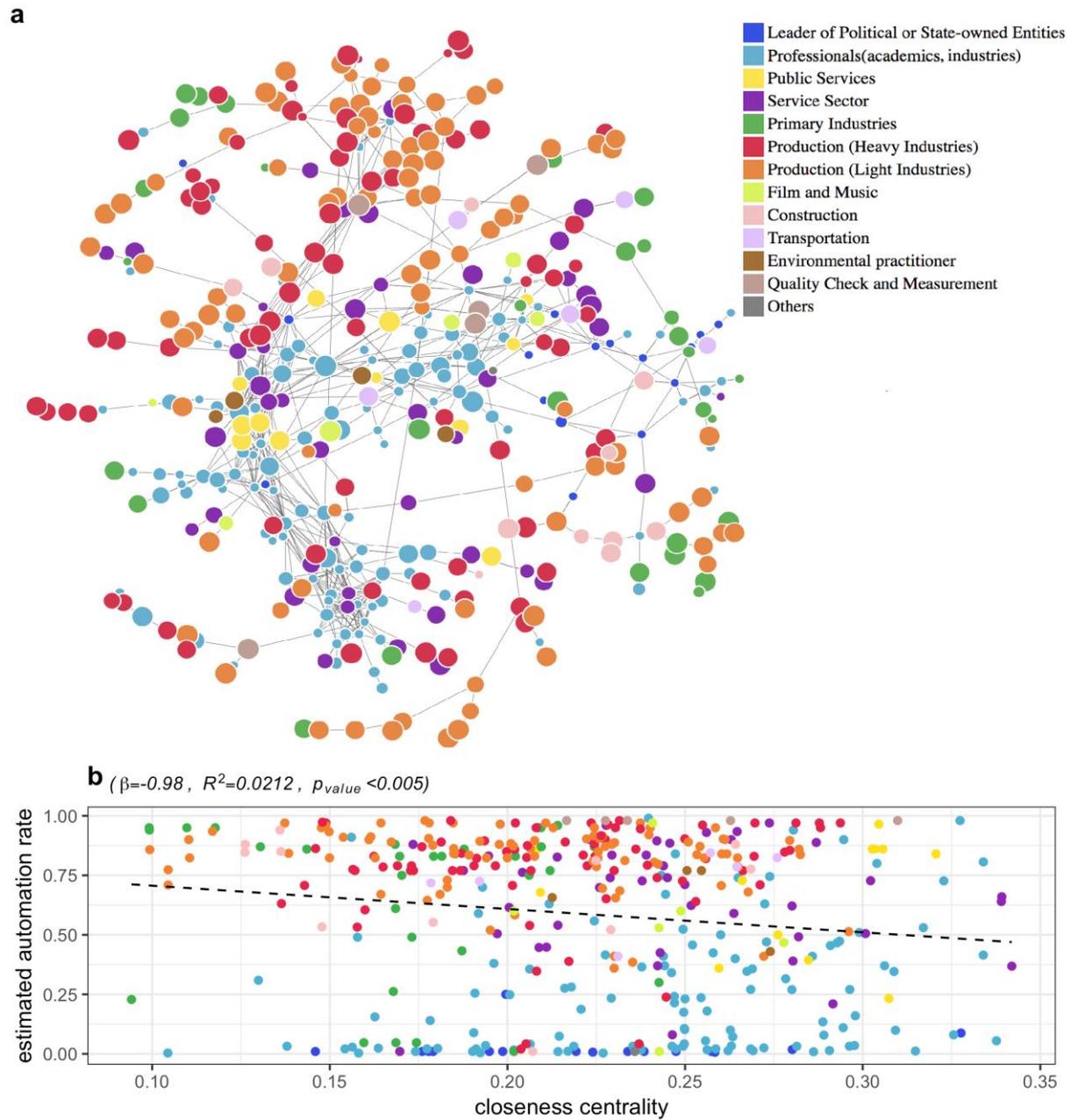

**Fig. 5. Evolution of positions in occupation space along with growing size.** (a), (b), (d), and (e) represent the positions of Putian, Beijing, Sanya, and Nanyang in the occupation space. (c) represents two distinct evolution paths of premium and non-premium cities in the occupation space. The result of the linear regression model is presented in the inset table, and city size is log-transformed.

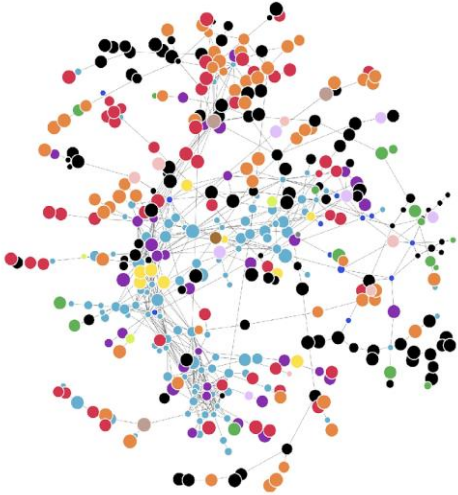
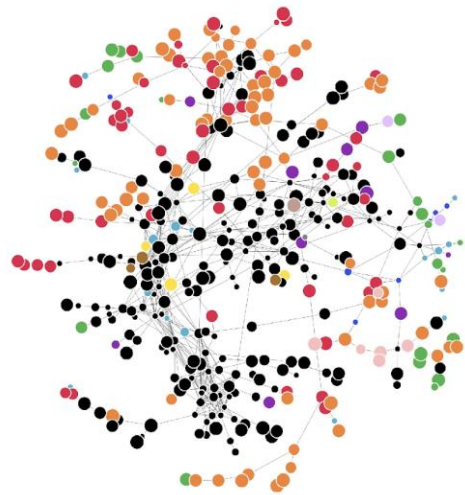
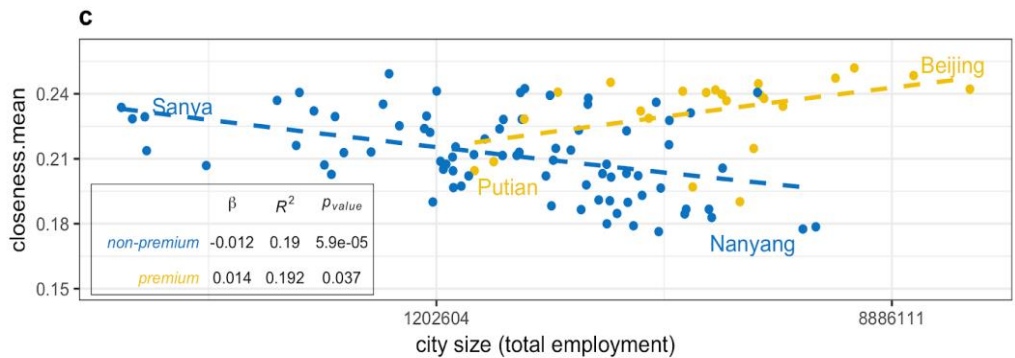
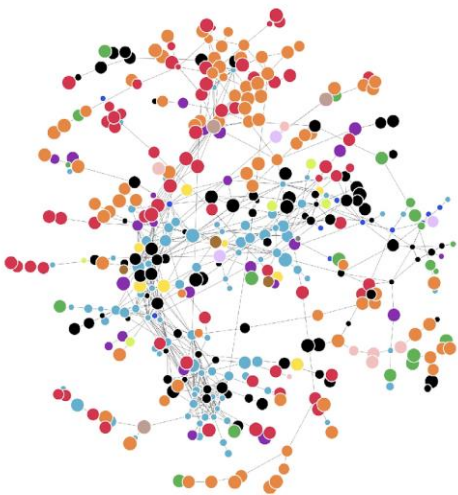
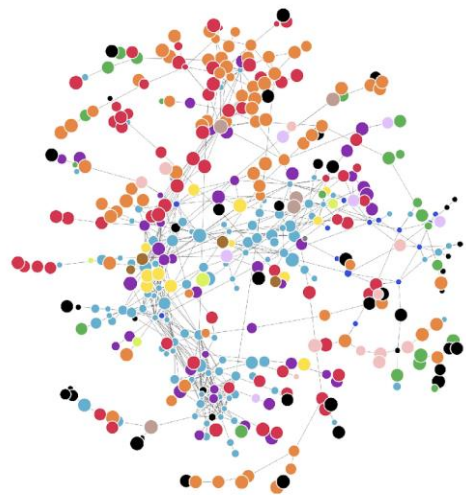

# Supplementary Materials

**Fig. S1. The fixed asset investments over city size.** Panel (a) shows result under the resources division and (b) the administrative division. The result of the linear regression is presented in tables, and the predictor city size is log-transformed.

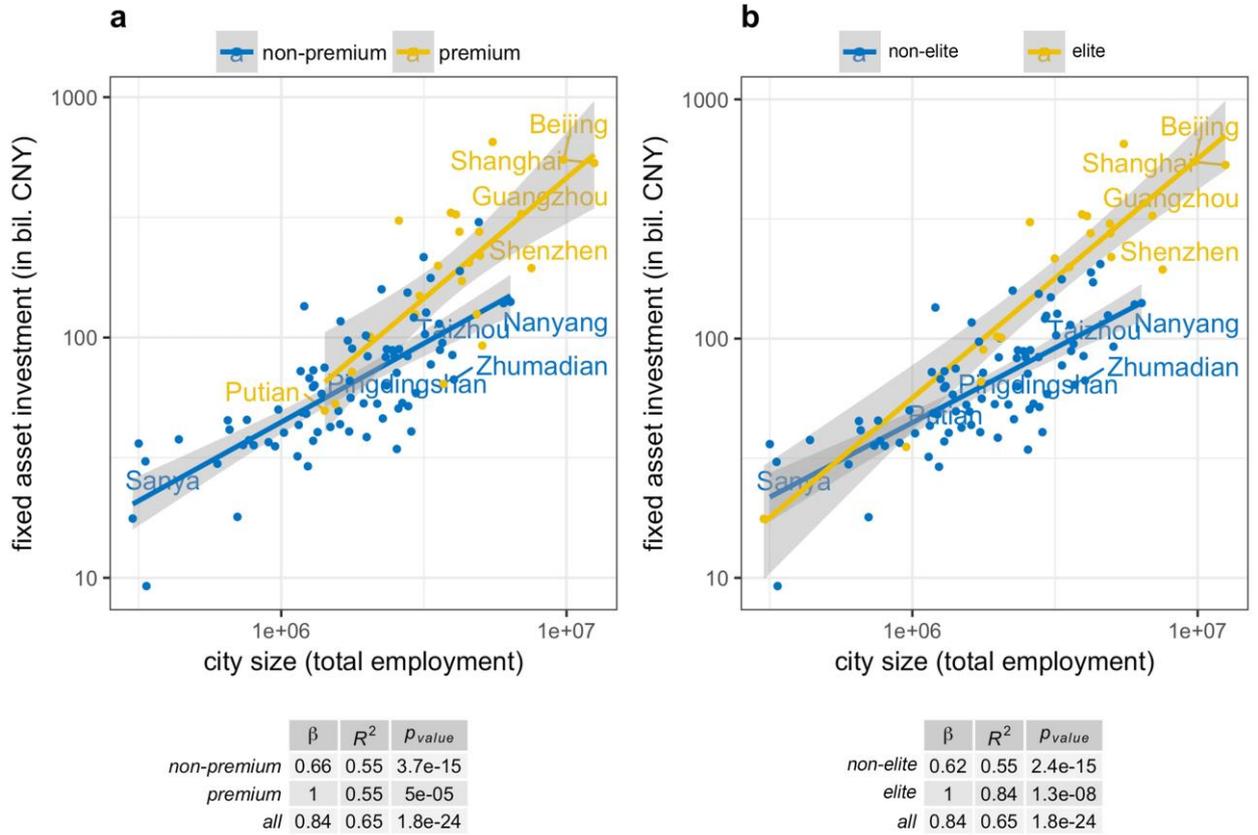

**Fig. S2. The comparison of expected job impact rate across Chinese and US cities.**
The predictor city size is log-transformed. The job impact rate and city size are obtained from (7).

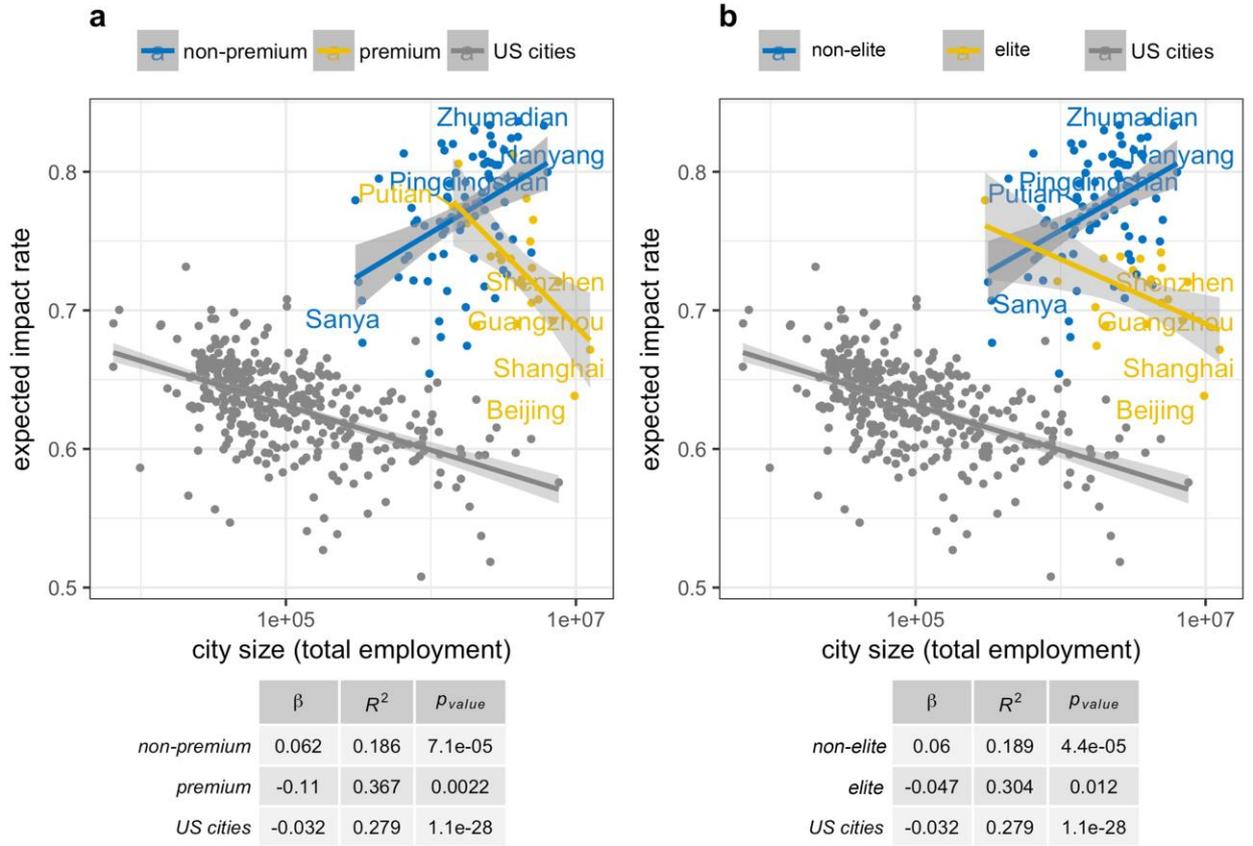

**Fig. S3. Industry diversity over the distance to the closest elite cities.** (a) The resources division, (b) The administrative division. The predictor distance to closest elite city is log-transformed.

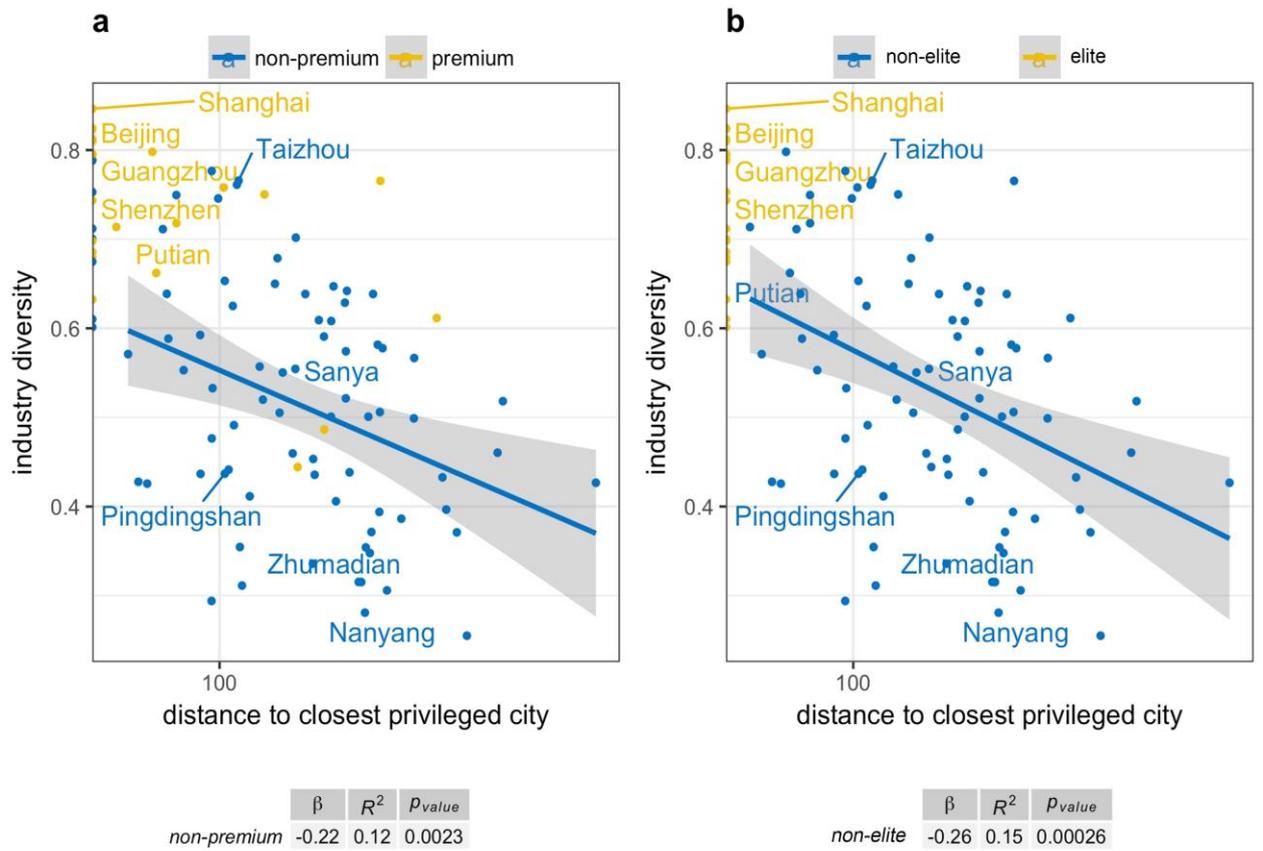

**Fig. S4. Population net gain over the distance to the closest elite cities.** (a) The resources division, (b) The administrative division. The predictor distance to closest elite city is log-transformed.

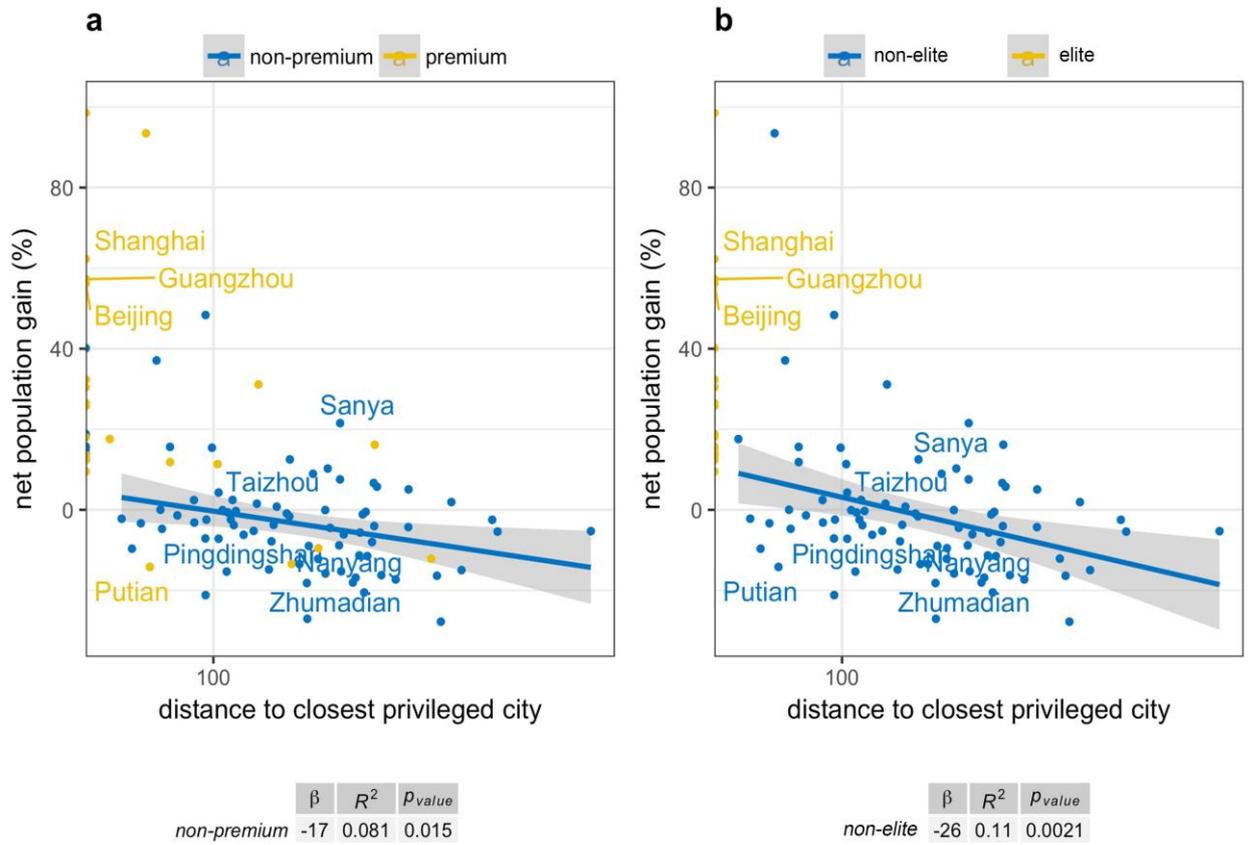

**Fig. S5. Job diversity over industry diversity.** (a) The resources division, (b) The administrative division.

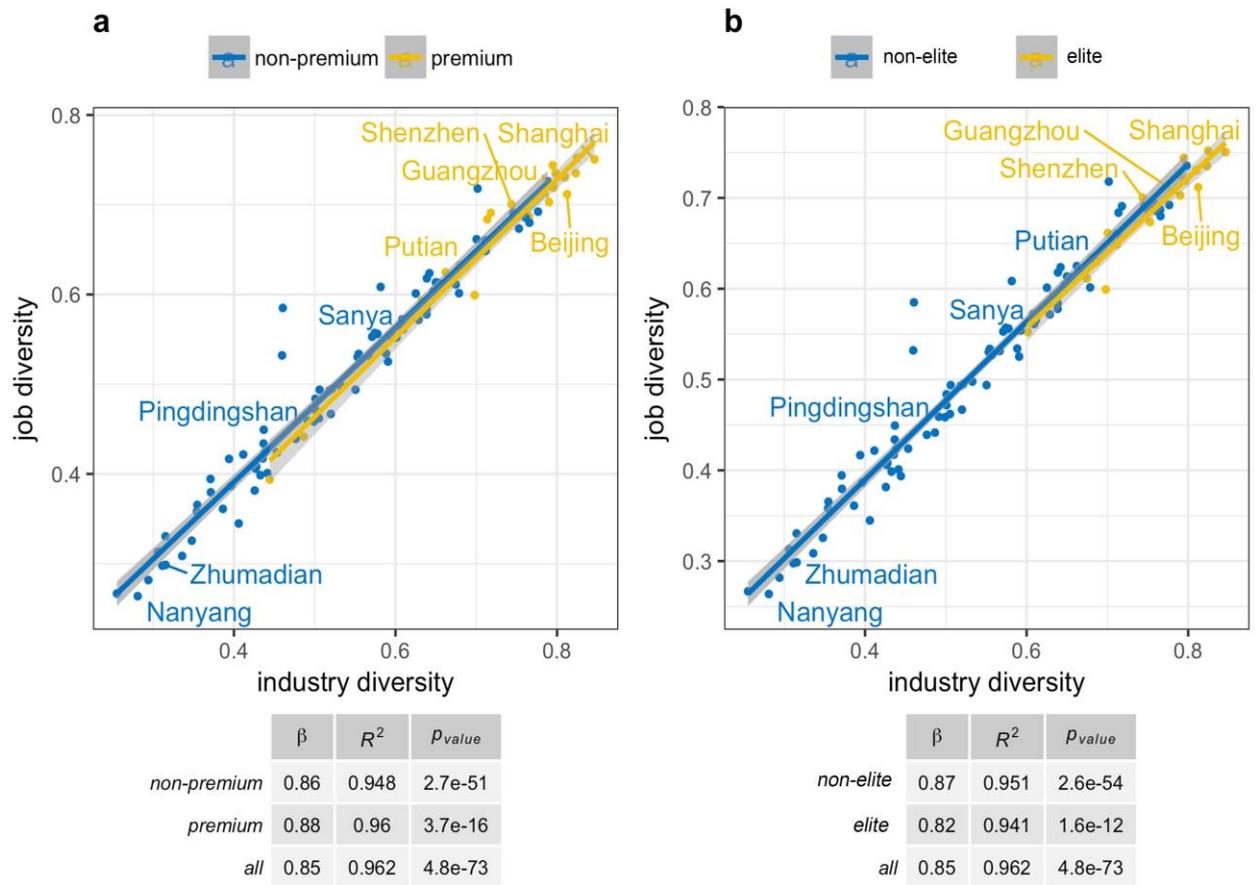

**Fig. S6. Job growth over city size between elite and non-elite cities.**

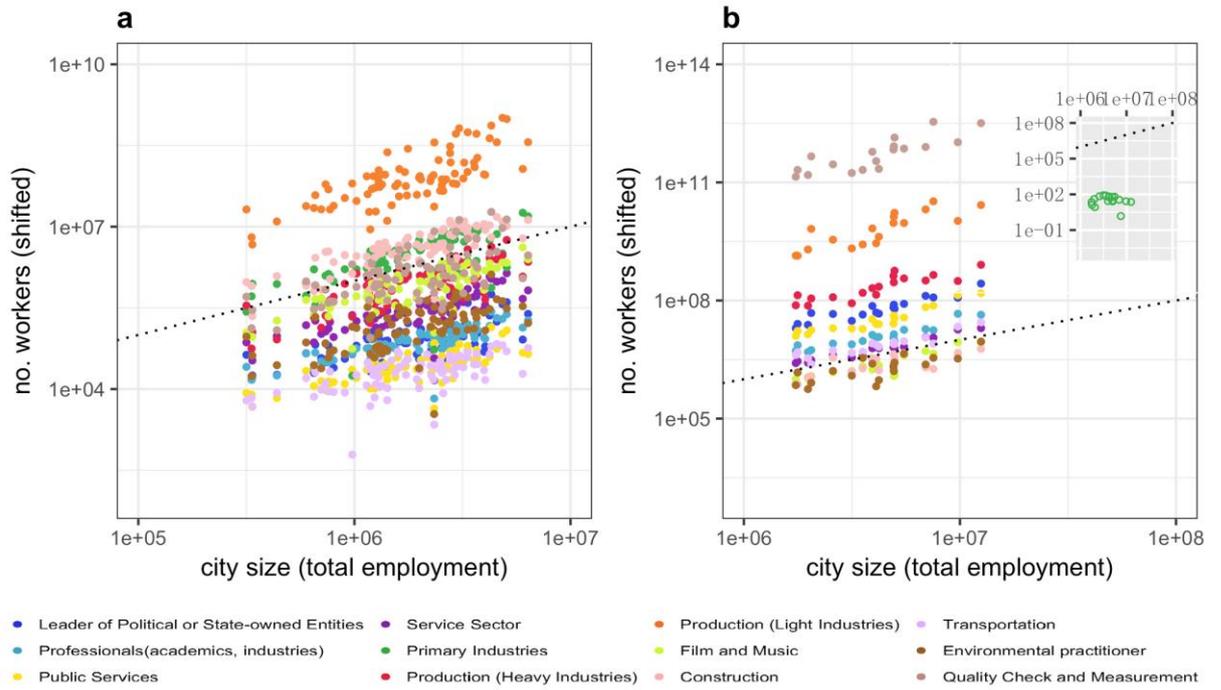

**Fig. S7. Allocations of vocational education resources over city size under the premium resources division.** (a) the vocational teachers grow superlinearly in premium cities and linearly in non-premium cities. (b) the vocational schools grow sublinearly in both premium and non-premium cities.

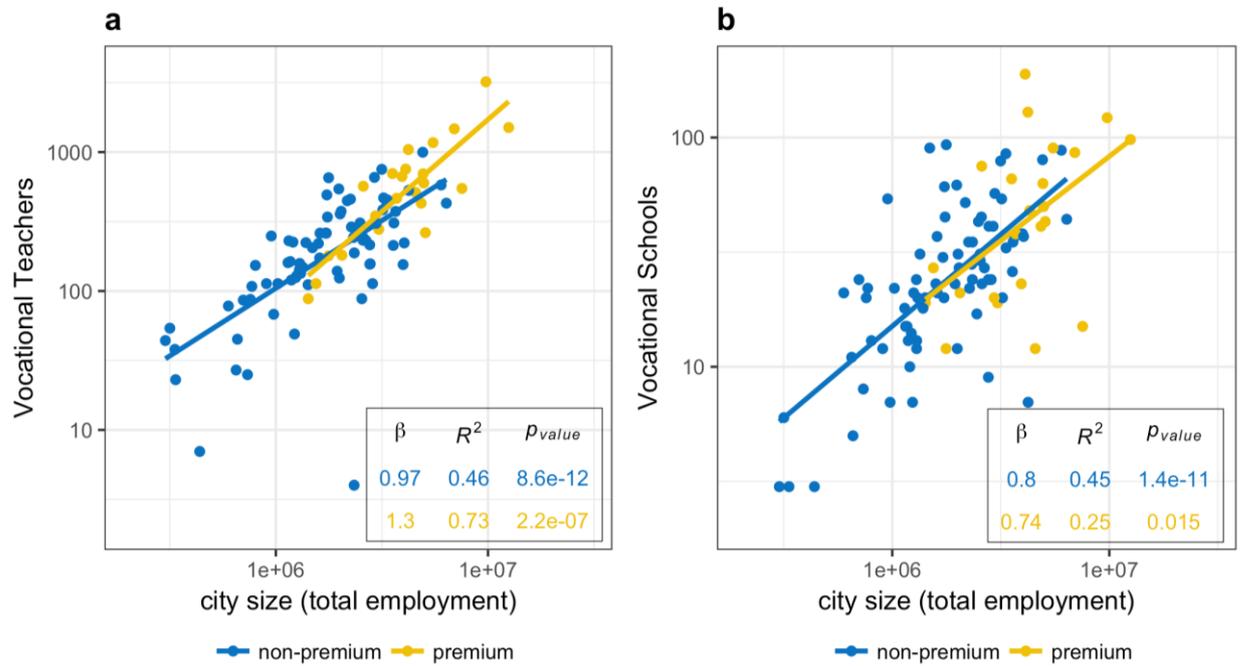

**Fig. S8. Allocations of vocational education resources over city size under the administrative division.** (a) the vocational teachers grow linearly in non-elite cities and sublinearly in elite cities. (b) the vocational schools grow sublinearly in both elite and non-elite cities.

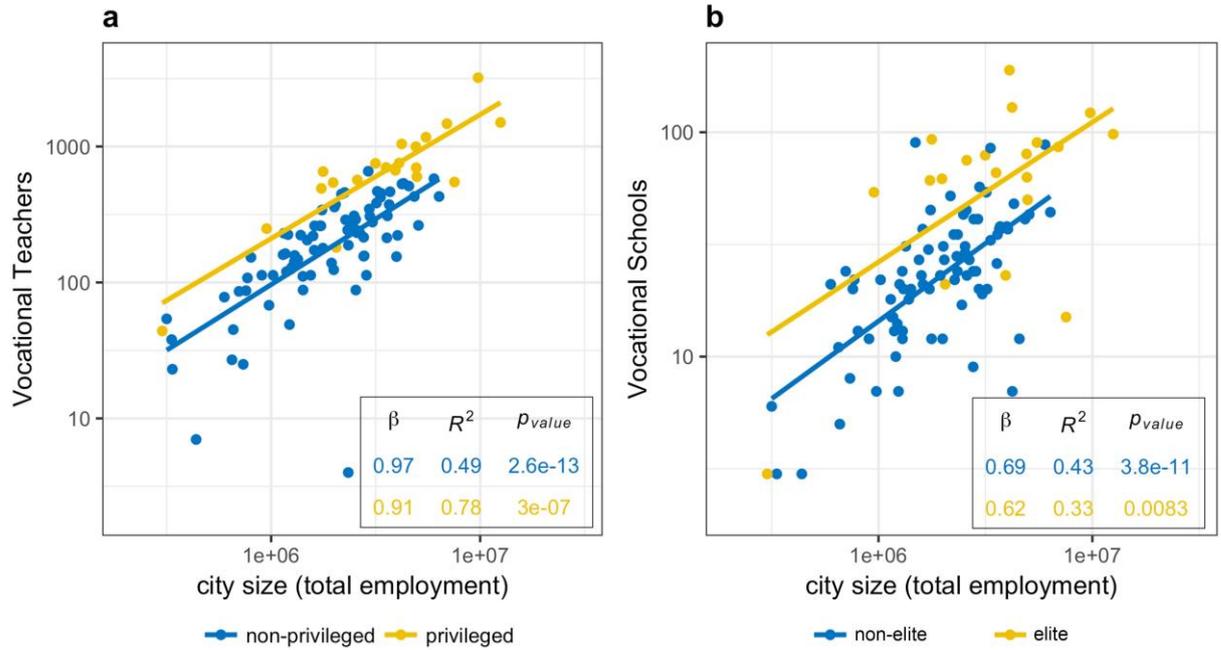

**Table S1. China's cities division system.** The numbers of universities under projects '211' and '985' and the daily operating frequency of bullet trains are shown in column 2, 3, 6, and 7. The estimation of job impact rate are shown in column 4 and 8.

| City | Universities Count | Bullet Trains | Expected Job Impact Rate | City | Universities Count | Bullet Trains | Expected job impact rate |
|---|---|---|---|---|---|---|---|
| Yibin | 0 | 0 | 81.99% | Puyang | 0 | 0 | 81.09% |
| Luzhou | 0 | 0 | 81.26% | Kaifeng | 0 | 58 | 82.60% |
| Lhasa[1] | 1 | 0 | 77.95% | Zhumadian | 0 | 93 | 83.65% |
| Heihe | 0 | 0 | 73.94% | Luohe | 0 | 97 | 79.15% |
| Beijing[1,2] | 35 | 349 | 63.83% | Pingdingshan | 0 | 0 | 79.52% |
| Shanghai[1,2] | 10 | 372 | 67.16% | Sanmenxia | 0 | 64 | 75.81% |
| Tianjin[1,2] | 3 | 261 | 70.78% | Zhengzhou[1,2] | 1 | 230 | 72.23% |
| Shenzhen[1,2] | 0 | 246 | 72.05% | Lanzhou[1] | 2 | 112 | 70.23% |
| Guangzhou[1,2] | 4 | 377 | 69.26% | Qingyang | 0 | 0 | 82.05% |
| Foshan[2] | 0 | 148 | 71.83% | Qingdao[1] | 1 | 67 | 74.17% |
| Zhuhai | 0 | 80 | 65.44% | Weihai | 0 | 46 | 76.16% |
| Huizhou | 0 | 131 | 77.24% | Jinan[1,2] | 1 | 248 | 73.73% |
| Jiangmen | 0 | 15 | 71.72% | Dongying | 0 | 0 | 75.48% |
| Kunming[1] | 0 | 73 | 72.91% | Liaocheng | 0 | 0 | 81.29% |
| Yuncheng | 0 | 43 | 75.35% | Linyi | 0 | 0 | 79.97% |
| Changzhi | 0 | 0 | 77.47% | Tai'an | 0 | 0 | 77.81% |

| | | | | | | |
|---|---|---|---|---|---|---|
| Linfen | 0 | 58 | 77.83% | Dezhou | 0 | 124 | 82.42% |
| Lvliang | 0 | 0 | 77.12% | Binzhou | 0 | 0 | 80.74% |
| Shuozhou | 0 | 0 | 76.28% | Chizhou | 0 | 53 | 77.40% |
| Jincheng | 0 | 0 | 76.40% | Bengbu[2] | 0 | 143 | 80.58% |
| Xinzhou | 0 | 0 | 79.92% | Suzhou | 0 | 88 | 78.97% |
| Yangquan | 0 | 39 | 72.39% | Tongling | 0 | 95 | 72.05% |
| Taiyuan[1] | 0 | 69 | 67.44% | Huainan | 0 | 45 | 73.86% |
| Jinzhong | 0 | 18 | 75.79% | Huaibei | 0 | 0 | 76.11% |
| Nanjing[1,2] | 8 | 335 | 69.02% | Huangshan | 0 | 49 | 72.15% |
| Yancheng | 0 | 0 | 79.67% | Liu'an | 0 | 71 | 80.49% |
| Xuzhou[2] | 1 | 273 | 78.07% | Xuancheng | 0 | 0 | 78.19% |
| Taizhou | 0 | 48 | 76.06% | Haozhou | 0 | 0 | 83.37% |
| Chifeng | 0 | 0 | 79.84% | Hefei[1,2] | 2 | 213 | 73.89% |
| Suizhou | 0 | 26 | 81.54% | Fangchenggang | 0 | 7 | 79.50% |
| Shiyan | 0 | 4 | 79.24% | Liuzhou | 0 | 128 | 77.70% |
| Shaoxing[2] | 0 | 168 | 74.06% | Guigang | 0 | 111 | 82.99% |
| Jiaxing[2] | 0 | 184 | 73.61% | Ningde | 0 | 56 | 71.40% |
| Huzhou[2] | 0 | 144 | 76.27% | Zhangzhou | 0 | 0 | 70.87% |
| Wenzhou[2] | 0 | 148 | 76.53% | Xiamen[1,2] | 1 | 212 | 68.91% |
| Zhoushan | 0 | 0 | 73.65% | Putian[2] | 0 | 183 | 77.49% |

| | | | | | | |
|---|---|---|---|---|---|---|
| Lishui | 0 | 63 | 69.21% | Longyan | 0 | 43 | 76.74% |
| Quzhou | 0 | 97 | 74.07% | Quanzhou[2] | 1 | 208 | 74.97% |
| Taizhou | 0 | 106 | 75.12% | Nanping | 0 | 75 | 76.41% |
| Jinhua | 0 | 46 | 72.58% | Xi'an[1,2] | 5 | 155 | 71.99% |
| Hangzhou[1,2] | 1 | 247 | 70.54% | Yan'an | 0 | 6 | 68.07% |
| Ningbo[1,2] | 1 | 163 | 73.07% | Baoji | 0 | 86 | 76.81% |
| Luoyang | 0 | 120 | 79.53% | Turpan | 0 | 31 | 67.66% |
| Shangqiu | 0 | 98 | 82.52% | Jilin | 0 | 94 | 78.46% |
| Xinyang | 0 | 122 | 81.58% | Liaoyuan | 0 | 0 | 81.32% |
| Xuchang | 0 | 68 | 80.56% | Songyuan | 0 | 0 | 82.01% |
| Nanyang | 0 | 4 | 83.33% | Hengyang[2] | 0 | 195 | 81.19% |
| Xinxiang | 0 | 75 | 80.48% | Xiangtan | 0 | 69 | 78.03% |
| Hebing | 0 | 60 | 76.50% | Haikou[1] | 0 | 57 | 72.10% |
| Anyang | 0 | 81 | 80.64% | Sanya | 0 | 51 | 70.71% |
| Jiaozuo | 0 | 14 | 78.18% | Guiyang[1] | 0 | 125 | 73.75% |

Note: Cities marked with superscript 1 and 2 are elite and premium cities respectively. It should be noticed that even though Lhasa isn't counted as a premium city due to the lack of bullet train passing-by, its railway investments are much higher than any other bullet train lines (*31*).